\documentclass{article}

\usepackage{arxiv}

\usepackage[utf8]{inputenc} % allow utf-8 input
\usepackage[T1]{fontenc}    % use 8-bit T1 fonts
\usepackage{hyperref}       % hyperlinks
\usepackage{url}            % simple URL typesetting
\usepackage{booktabs}       % professional-quality tables
\usepackage{amsfonts}       % blackboard math symbols
\usepackage{nicefrac}       % compact symbols for 1/2, etc.
\usepackage{microtype}      % microtypography
\usepackage{lipsum}		% Can be removed after putting your text content
\usepackage{graphicx}
\usepackage{doi}

\title{The cosmological constant emerging from a symmetry invariant}

\author{ \href{https://orcid.org/0000-0002-2509-5048}{\includegraphics[scale=0.06]{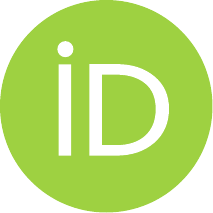}\hspace{1mm}Ivan Arraut}\\
	 Institute of Science and Environment and FBL,\\
  University of Saint Joseph,\\
Estrada Marginal da Ilha Verde, 14-17, Macao, China\\
	\texttt{ivan.arraut@usj.edu.mo} \\
}

\renewcommand{\shorttitle}{\textit{arXiv} Template}

\hypersetup{
pdftitle={A template for the arxiv style},
pdfsubject={q-bio.NC, q-bio.QM},
pdfauthor={Ivan Arraut},
pdfkeywords={Dark Energy, Zero point quantum fluctuations, Cosmological constant},
}

\begin{document}
\maketitle

\begin{abstract}
The cosmological constant is normally introduced as an additional term entering the Einstein-Hilbert (EH) action. In this letter we demonstrate that instead, it appears naturally from the standard EH action as an invariant term emerging from spacetime symmetries. We then demonstrate that the same constraint emerging from this invariant, suppresses the short wavelength modes and it favors the long wavelength ones. In this way, inside the proposed formulation, the observed value for the vacuum energy density is obtained naturally from the zero-point quantum fluctuations.
\end{abstract}

% keywords can be removed
\keywords{Dark Energy \and Zero point quantum fluctuations \and Cosmological constant}

\section{Introduction}

The cosmological constant is normally introduced by hand inside the Einstein-Hilbert action in order to reproduce the accelerated expansion of the universe \cite{1, 2, 3, CCproblem, CCproblem2, 4, 7, 8, 11, 12, 13}. Its theoretical value was calculated in the past by integrating the zero-point quantum fluctuations over all the possible modes, using as a cut-off the Planck length. This calculation generates a huge value for the vacuum energy density, so big that its mismatch with the observations corresponds to 120 orders of magnitude \cite{3, CCproblem, CCproblem2, 5}. This discrepancy between the theoretical and experimental value, is normally dubbed as "the worst theoretical prediction in physics", and for obvious reasons. As a consequence of this, several modifications of gravity have emerged, trying to suggest possible solutions to the problem. Among all the possibilities, we have non-local formulations of gravity, scalar-tensor theories, among others \cite{9, 10}. The problem of modifying gravity, which implies a modification of the Einstein-Hilbert action, is that several problems like instabilities, ghosts, among other pathologies emerge during the process. Additionally, the introduction of new fields and/or degrees of freedom requires some consistent physical explanation, difficult to justify in standard scenarios. In this paper, without modifying gravity, we demonstrate that the cosmological constant emerges naturally, inside the scenario of General Relativity, from a symmetry invariant valid at cosmological scales. The invariant term then appears reproducing the appropriate potential (de-Sitter) function inside the equations describing the motion of a test particle. If we use the constraint emerging from the symmetry invariant, in order to calculate the vacuum energy density from the zero-point quantum fluctuations, we then find a natural suppression of the short wavelength modes. This then generate a scenario in favor of the the long wavelength modes. The final result shows that the correct value for the vacuum energy density, exactly $120$ order of magnitude smaller than the standard theoretical value, is obtained from this formulation.

\section{Standard gravity with a positive Cosmological constant}   \label{s2}

The Einstein equations can be expressed as \cite{5}

\begin{equation}   \label{Einstein1}
R_{\mu\nu}-\frac{1}{2}g_{\mu\nu}R-\Lambda g_{\mu\nu}=8\pi G T_{\mu\nu}.
\end{equation}
Here $R_{\mu\nu}$ is the Ricci tensor and $R=g_{\mu\nu}R^{\mu\nu}$ is the curvature scalar. Additionally, $G$ is the Newtonian constant, $T_{\mu\nu}$ is the energy-momentum tensor (source term), $g_{\mu\nu}$ is the metric and $\Lambda$ is the cosmological constant. From the solutions of the equation (\ref{Einstein1}), we can explain the accelerated expansion of the universe \cite{5, 7}. Eq. (\ref{Einstein1}) can be obtained from the Einstein-Hilbert (EH) action expressed as follows

\begin{equation}   \label{Action}
S=\frac{1}{2\kappa}\int d^4x\sqrt{-g}(R-2\Lambda)+S_M.
\end{equation}
Here $S_M$ is the matter action related to the source term $T_{\mu\nu}$ in eq. (\ref{Einstein1}). In addition, $\kappa=8\pi G$  and $g$ is the determinant of the metric \cite{5}.

\subsection{The spherically symmetric solution in General Relativity with a positive cosmological constant}

The standard Schwarzschild de-Sitter metric is defined as \cite{7, 8}

\begin{eqnarray}   \label{SDSmetric}
ds^2=-\left(1-\frac{2GM}{r}-\frac{1}{3}\Lambda\right)dt^2\\\nonumber
+\left(1-\frac{2GM}{r}-\frac{1}{3}\Lambda\right)^{-1}dr^2+r^2d\Omega^2.
\end{eqnarray}
From this metric, we can derive the equations of motion of a test particle moving around a gravitational source. This calculation has been done previously in \cite{Marek} and it comes from expanding the equation $g_{\mu\nu}U^\mu U^\nu=-\epsilon$. Here $U^\mu=dx^\mu/d\lambda$, with $\lambda$ being the affine parameter which for massive test particles is just the proper time. After doing the corresponding expansion by using the metric (\ref{SDSmetric}), we get

\begin{equation}   \label{varepsilon}
\frac{1}{2}\left(\frac{dr}{d\lambda}\right)^2+V(r)=\varepsilon,    
\end{equation}
with the potential defined as 

\begin{equation}
V(r)=-\frac{GM}{r}-\frac{1}{6}\Lambda r^2+\frac{L^2}{2r^2}-\frac{GML^2}{r^3}.    
\end{equation}
In eq. (\ref{varepsilon}), $\varepsilon$ is just a constant of motion depending on the cosmological constant, the angular momentum and on the "conserved" energy of the system \cite{Marek}. We have to remark however, that the energy is not conserved in a universe in expansion. This will be an important point to consider in the coming calculations.

\section{The cosmological constant emerging as a consequence of a symmetry invariant quantity}

If we set $\Lambda=0$ in eq. (\ref{Action}), we then  have the standard Einstein-Hilbert action. If we are analyzing cosmological scales, then the energy is not a conserved quantity anymore and then we cannot consider the time-component of the Killing vector for finding a conserved quantity as it is usually done for the analysis of the motion of a test particles \cite{5, Marek, Ivan}. Additionally, since the spacetime expansion dominates at such scales, then local variations of the radial coordinate can be ignored in relation to the changes occurring over the time-coordinate ($dt/d\tau>>dr/d\tau$). Then we can develop the general equation for the motion of a test particle as follows

\begin{equation}   \label{Uptown}
-\left(1-\frac{2GM}{r}\right)\left(\frac{dt}{d\lambda}\right)^2+Cr^n\left(\frac{d\phi}{d\lambda}\right)^2\approx-\epsilon. 
\end{equation}
Here $n$ is the power of the radial coordinate which will be fixed subsequently based on the symmetry constraint related to the angular coordinate. $C$ on the other hand, is a parameter which is part of the angular component of the metric. For simplicity, we will consider a massive test particle and the $\epsilon=1$ \cite{5}. The equation of motion emerging from eq. (\ref{Uptown}) is 

\begin{equation}   \label{Uptown2}
\frac{1}{2}\left(\frac{dt}{d\lambda}\right)^2-\frac{Cr^{n+1}}{2(r-2GM)}\left(\frac{d\phi}{d\lambda}\right)^2\approx\frac{r}{r-2GM}.    
\end{equation}
This is then an equation for the time-coordinate instead of the radial coordinate because at cosmological scales all the universe expansion must be attributed to changes on the time coordinate. At the same scales $r>>GM$ and then eq. (\ref{Uptown2}) becomes

\begin{equation}   \label{Uptown3}
\frac{1}{2}\left(\frac{dt}{d\lambda}\right)^2-C\frac{r^n}{2}\left(\frac{d\phi}{d\lambda}\right)^2\approx 1.    
\end{equation}
The Killing vector related to the angular coordinate is $K^\mu=(0, 0, 0, 1)$ and then the one-form defining the conserved quantity is

\begin{equation}   \label{omegagaga}
\beta=Cr^n\left(\frac{d\phi}{d\lambda}\right)=\frac{C}{r^2}\left(\frac{d\phi}{d\lambda}\right).    
\end{equation}
Then $n=-2$ is the value taken by the conserved quantity for the cosmological case. The consistency of this value will be perceived in the coming results.
If we replace the result (\ref{omegagaga}) inside eq. (\ref{Uptown3}), then we get

\begin{equation}   \label{Uptown4}
\frac{1}{2}\left(\frac{dt}{d\lambda}\right)^2-\frac{\Lambda}{3}r^2\approx 1,
\end{equation}
where we have defined $\beta^2/C=\Lambda/3$, where $\beta$ is the square of the new conserved quantity related to the angle of rotation described by the test particle when it moves around the source. It is important to remark that the existence of the new invariant, requires the angular portion of the metric to change as $r^2d\Omega^2\to \frac{C}{r^2}d\Omega^2$. The freedom provided by the vacuum solution for the Schwarzschild-like metric, allows us to derive this solution without challenging any principle of GR. Then as it can be seen, a cosmological constant-like term emerges from symmetry arguments. This however does not guarantee that we have solved the dark energy problem. The solid proof showing that the conservation of $\beta$ marks the solution of the dark energy problem, comes from the calculation of the vacuum energy density from the zero-point quantum fluctuations of the vacuum. In a moment we will see that the constraint coming from the new conserved quantity (\ref{omegagaga}), defined at cosmological scales, is the key ingredient for solving the puzzle.

\section{Vacuum energy: Zero point quantum fluctuations}

The standard calculation (without any constraint coming from symmetry) for the vacuum energy density comes from the following expression

\begin{equation}   \label{Vac}
E_0^S\approx\frac{1}{(2\pi)^3}\int_0^{k_{max}}\hbar\omega d^3k.    
\end{equation}
Here $k_{max}$ is taken as the Planck scale $k_{max}\sim 1/l{pl}$. The expression (\ref{Vac}) is the energy coming from the ground state of a quantum field. The final result is $E_0^S\sim \hbar k_{max}^4=\hbar/l_{pl}^4$, where the index $S$ means standard (calculation). In this paper however, we have noticed that there is a constraint emerging from the symmetry invariant $\beta$ in eq. (\ref{omegagaga}). This constraint has to be considered at the moment of doing the integration in eq. (\ref{Vac}) as follows

\begin{equation}   \label{Vac2}
E_0\approx\frac{1}{(2\pi)^3}\int_0^{k_{max}}\int_0^{\phi_{max}}\int_{-\pi}^{\pi}\hbar k^3 sin\theta dk d\phi d\theta.
\end{equation}
This is a triple integral with the volume element being $k^2 sin\theta dk d\phi d\theta$. Additionally, we take $\omega\sim k$ (taking the speed of light as $c=1$). The integration over $\theta$ just gives us a factor $2$. The important integration is the one done over $\phi$ because it is constrained by the invariant defined in eq. (\ref{omegagaga}). If we integrate eq. (\ref{omegagaga}), by keeping fixed all the other quantities, then we get the maximum angle $\phi$ as $\phi_{max}=r^2\beta\lambda/C\sim \frac{\beta\lambda}{Ck^2}$, considering the uncertainty principle condition $\Delta r\Delta k\sim 1$ and then the corresponding condition $r\sim 1/k$. If we replace $\phi_{max}$ inside the integral (\ref{Vac2}), then we get

\begin{equation}
E_0\approx\frac{2}{(2\pi)^3}\int_0^{k_{max}}\int_0^{\frac{\beta\lambda}{Ck^2}}\hbar k^3 d\phi dk.
\end{equation}
We now proceed to evaluate this double integral. If we integrate first the integration over the variable $\phi$ and subsequently over $k$, then we get

\begin{equation}
E_0\sim \frac{\hbar}{(2\pi)^3}\frac{\beta\lambda}{C}k_{max}^2\sim \frac{\hbar}{(2\pi)^3}\frac{\lambda}{C\beta}k_{max}^2\Lambda.
\end{equation}
If we consider the UV cut-off as the Planck scale, then $k_{max}\sim 1/l_{pl}$. In addition, $\Lambda$ is a free parameter defined as the IR cut-off $\Lambda\sim 1/r_{\Lambda}^2$, in agreement with the notation used in \cite{8, Marek}. Here $r_\Lambda\sim 10^{26}mt$, namely, the size of the observable universe. Then the quotient between the standard value $E_0^S$ and the value $E_0$ which is constrained by symmetry is $E_0^S/E_0\sim 10^{120}$, namely, exactly the $120$ orders of magnitude as it should be. This result solves the Dark Energy problem, demonstrating that this form of energy really comes from the zero-point quantum fluctuations. This also demonstrates that the standard formulation of Quantum Field Theory is correct \cite{QFT}. It is important to remark that the combination of parameters $\lambda$, $\beta$ and $C$, satisfy the condition $\frac{\lambda}{\beta C}\sim1$, with $\lambda$ being a representative scale for the proper time perceived by a particle when it moves through spacetime.

\section{Conclusions}

In this paper we have demonstrated that the cosmological constant term appears naturally the standard theory of General Relativity as a consequence of symmetry arguments. More specifically, this term (Cosmological Constant) emerges as a parameter related to an invariant of the theory, valid at cosmological scales. Interestingly, the constraint emerging from the symmetry-invariant, favors the long wavelength modes and suppresses the short wavelength ones. In this way, the correct value for the vacuum energy density, which is exactly $120$ orders of magnitude smaller than the value obtained by using the standard calculation, emerges from this formulation. The results in this paper also demonstrate that the cosmological constant is fully related to the energy coming from the zero-point quantum fluctuations, something which is strongly supported by the experimental evidence and from the standard formulation of Quantum Field Theory \cite{QFT}.

%\bibliographystyle{unsrtnat}
%\bibliography{references}  %%% Uncomment this line and comment out the ``thebibliography'' section below to use the external .bib file (using bibtex) .

%%% Uncomment this section and comment out the \bibliography{references} line above to use inline references.

\end{document}